\documentclass{article}

\usepackage[preprint]{neurips_2024}
\usepackage{graphicx}
\usepackage[dvipsnames]{xcolor}
\usepackage[utf8]{inputenc} 
\usepackage[T1]{fontenc}    
\usepackage{hyperref}       
\usepackage{url}            
\usepackage{booktabs}       
\usepackage{amsfonts}       
\usepackage{nicefrac}       
\usepackage{microtype}      
\usepackage{xcolor}         
\usepackage{float}
\usepackage[raster,most]{tcolorbox}
\usepackage{natbib}
\setcitestyle{numbers,square}
\usepackage{enumitem}
\usepackage{multirow}
\usepackage{multicol}

\title{Tibetan-TTS: Low-Resource Tibetan Speech Synthesis with Large Model Adaptation}

\definecolor{gray}{HTML}{777777}

\author{
	Jiaxu He$^1$, Chao Wang$^3$, Jie Lian$^1$, Yuqing Cai$^4$, \\ \textbf{Yongxiang Li$^1$, Renzeg Duojie$^2$, Jie Li$^{1,*}$} \\
	\\
	Xingchen AGI Lab, China Telecom Artificial Intelligence Technology Co. Ltd\\
     \\ {Xizang University, Lhasa, China}  \\
     \\ {Qinghai Normal University, Xining, China}  \\
      \\ {University of Electronic Science and Technology of China,Chengdu,China}  \\
	\\
	\texttt{lij86@chinatelecom.cn}\\
}

\begin{document}

	\maketitle
	\renewcommand{\thefootnote}{}
	\footnotetext{$^*$ Corresponding author.}
	\setcounter{footnote}{0}
	
	\begin{abstract}
   
    Tibetan text-to-speech (TTS) has long been challenged by scarce speech resources, significant dialectal variation, and the complex mapping between written text and spoken pronunciation. To address these issues, this work presents, to the best of our knowledge, the first large-model-based Tibetan TTS system in the industry, built upon a large speech synthesis model developed by Xingchen AGI Lab. The proposed system integrates data quality enhancement, Tibetan-oriented text representation and tokenizer adaptation, and cross-lingual adaptive training for low-resource Tibetan speech synthesis. Experimental results show that the system can generate stable, natural, and intelligible Tibetan speech under low-resource conditions. In subjective evaluation, the MOS scores of the syllable-level and BPE-based systems reach 4.28 and 4.35, while their pronunciation accuracies reach 97.6\% and 96.6\%, respectively, outperforming an external commercial Tibetan TTS interface. These results demonstrate that combining a large-model backbone with Tibetan-oriented text representation adaptation and cross-lingual adaptive training enables highly usable low-resource Tibetan speech synthesis, and also provides a technical foundation for future unified multi-dialect Tibetan speech synthesis.
		
	\end{abstract}
	
	\section{Introduction}

    Tibetan is one of the most important minority languages in China, carrying a long history and rich cultural heritage. However, due to the long-standing scarcity of linguistic resources, significant dialectal variation, and the complex mapping between the conservative written system and modern spoken forms, the development of Tibetan text-to-speech (TTS) technology has lagged far behind that of high-resource languages such as Mandarin Chinese and English.In domains including public services, education, and cultural heritage preservation, high-quality Tibetan speech synthesis holds substantial social value, serving as a key enabling technology for information accessibility, the protection of endangered dialects, and the digital dissemination of ethnic culture.\cite{huang2025tibetanAI}
    
    At present, building a practical Tibetan TTS system capable of covering the major dialect groups---\"{U}-Tsang, Amdo, and Kham---has become an urgent requirement for improving technological usability and service coverage. Achieving this goal, however, faces several fundamental challenges. First, the Tibetan writing system remains relatively conservative, while modern spoken Tibetan has undergone substantial phonological evolution across different dialects. This results in systematic and non-bijective mappings between text and speech, greatly increasing the complexity of automatic text-to-phoneme conversion. Second, Tibetan exhibits pronounced internal dialectal diversity: \"{U}-Tsang dialects are tonal, Amdo dialects are non-tonal but feature complex consonant clusters, and Kham dialects share characteristics of both. These substantial phonological differences make it extremely difficult to construct a unified speech synthesis model.\cite{gao2025tluetibetanlanguageunderstanding} In addition, the overall ecosystem for Tibetan language technology remains underdeveloped. Large-scale, high-quality TTS datasets are scarce, and mature natural language processing tools---such as Tibetan word segmentation and syntactic analysis---are still limited, which further constrains the accuracy of front-end text processing and increases the cost and risk of system development.\cite{gao2025tibstccotmultidomaininstructiondataset,huang2025tibstclargescalestructuredtibetan}
    
    To address the above challenges, this paper investigates the construction of a Tibetan text-to-speech (TTS) system under low-resource conditions. Building on our collaboration with Xizang University, we make full use of its dialectal speech resources and linguistic expertise to explore effective modeling strategies for non-ideal data conditions. Within a large-model-based speech synthesis framework, this work focuses on three key aspects: data processing, text representation adaptation, and cross-lingual transfer modeling, and proposes a technical route for low-resource Tibetan TTS that integrates large-model capability with domain knowledge. The main contributions of this work are as follows:

    \begin{itemize}
        \item
        First, we propose a unified data processing pipeline for low-quality, multi-source Tibetan speech data. To address issues such as noise interference, inconsistent annotations, and irregular text forms in raw data, we develop a systematic data processing scheme covering audio cleaning, text normalization, and speech-text alignment. This pipeline enables the transformation of raw recordings into high-quality training data and provides a reliable foundation for subsequent model training.
        \item
        Second, we propose a low-resource text representation modeling method that combines domain knowledge with large-model adaptation. To address the mismatch between the complex structure of Tibetan writing and the default tokenization strategy of the pretrained model, we design a Tibetan-oriented tokenization strategy at the text representation level and integrate it into the large model's input representation module. This allows the model to learn using units that are more consistent with Tibetan linguistic characteristics, thereby improving the stability and accuracy of text-to-speech mapping under limited data conditions.
        \item
        Third, we propose a cross-lingual adaptive training method for low-resource scenarios.By leveraging the cross-lingual representation capability of the pretrained large model and combining it with lightweight fine-tuning strategies, the proposed method enables effective transfer of speech synthesis capability to Tibetan under limited supervision. Benefiting from the general modeling capability of the pretrained backbone, the system achieves more stable training and synthesis under low-resource conditions, while improving the naturalness and intelligibility of the target dialect speech.
    \end{itemize}

     In summary, this work establishes, to the best of our knowledge, the first large-model-based Tibetan TTS framework in the industry for low-resource scenarios, integrating data processing, text representation adaptation, and cross-lingual adaptive training into a unified technical solution. It helps address long-standing resource and engineering bottlenecks in Tibetan speech synthesis, while also providing practical insights for TTS development in other low-resource languages.

	\section{Related Work and Technical Approaches}
	\label{sec:method}

    The development of Tibetan speech synthesis has mainly focused on three major challenges: low-resource modeling, complex text-to-speech mapping, and significant dialectal variation. Early studies largely adopted mainstream neural speech synthesis architectures such as Tacotron2 and FastSpeech2\cite{shen2018tacotron2,ren2020fastspeech2}, which demonstrated the feasibility of deep learning methods for Tibetan TTS\cite{zhao2019lhasa,zhou2024tibetan}. These works typically concentrated on improving the stability of alignment between textual and acoustic features, alleviating problems such as skipped or repeated pronunciations and monotonous prosody through explicit duration modeling, improved attention mechanisms, or enhanced prosody modeling. However, due to the limited scale of training data and the substantial differences among Tibetan dialects, approaches that rely entirely on training from scratch often struggle to achieve stable performance.

    With the development of cross-lingual transfer learning, leveraging knowledge from pretrained models in high-resource languages to support low-resource language modeling has gradually become an important direction for improving Tibetan speech synthesis. Existing studies suggest that, compared with training solely on small-scale target-language data, fine-tuning cross-lingually pretrained models can more effectively mitigate the instability caused by data scarcity and improve the usability of the resulting system\cite{researchTibetanRhythmic,TMDTTS2025,MSDTTS2025}.

    In recent years, the rapid development of large-scale speech and language models has driven speech synthesis toward a new paradigm represented by two-stage architectures\cite{cosyvoice2,vall-e,peng2026vibevoice}. In this framework, the input text is first mapped into high-level semantic or intermediate acoustic representations by a large language model or speech-language model, and is then converted into the final waveform by a neural vocoder or an efficient acoustic decoder\cite{kong2020hifigangenerativeadversarialnetworks,DBLP}. Compared with traditional single-stage models, two-stage architectures have demonstrated better scalability and transferability in multilingual, multi-speaker, and style-controlled speech synthesis tasks. Their key advantage lies in the ability to learn relatively universal speech representation structures from large-scale multilingual pretraining data. As a result, when adapted to low-resource languages, these models can achieve promising synthesis quality with only limited supervised data and relatively small-scale parameter updates\cite{YourTTS, sft}.

    From the perspective of technical selection, Tibetan speech synthesis requires a careful trade-off between a traditional highly modular front-end and a fully end-to-end modeling approach. The former offers stronger controllability, but depends heavily on linguistic rules and expert knowledge\cite{Ying_2024, g2pM}. The latter reduces the burden of front-end design, but under low-resource conditions it often suffers from training instability and frequent pronunciation errors. Therefore, a hybrid solution that combines pretrained large models with lightweight front-end adaptation becomes a reasonable choice for balancing controllability and modeling performance.

    Furthermore, within current large-model-based speech synthesis systems, multiple representative technical routes have emerged around the two-stage generation paradigm. Autoregressive speech language models generally perform well in high-level semantic modeling, consistency control, and naturalness, but their step-by-step generation process leads to relatively high inference costs and is more prone to accumulated errors such as repetition and omission in low-resource scenarios. Diffusion-based or other continuous acoustic generation models may offer advantages in synthesis quality, but they typically require longer sampling processes and impose higher demands on training data scale and engineering complexity. In contrast, a two-stage framework that employs an autoregressive language model for high-level semantic modeling and Flow Matching for acoustic generation can achieve a better balance among semantic consistency, generation efficiency, speech quality, and training stability. In addition, such frameworks usually provide relatively mature pretraining and fine-tuning paradigms, making them a more suitable backbone solution for low-resource Tibetan speech synthesis\cite{e2tts2024,chen2025f5ttsfairytalerfakesfluent,popov2021gradtts}.
    
    Based on the above considerations, this work adopts a large speech synthesis model developed by Xingchen AGI
    Lab as the backbone model. The model follows a two-stage speech generation architecture that combines an AR language model with Flow Matching, and is pretrained on approximately 200,000 hours of Chinese-English mixed speech data and 5,000 hours of multi-dialect speech data, endowing it with strong cross-lingual representation and generation capabilities. On this basis, this work focuses on data cleaning, text representation adaptation, and lightweight fine-tuning methods for Tibetan speech synthesis, with the goal of improving adaptation efficiency and synthesis quality under limited-resource conditions.

	\section{System Design and Realization}

	\subsection{A Unified Quality Enhancement Pipeline for Low-Resource, Multi-Source Tibetan Speech Data}
	



    High-quality and strictly aligned speech-text parallel data are a prerequisite for effective training and stable fine-tuning of low-resource speech synthesis models. However, in real-world Tibetan speech scenarios, the raw data often exhibit substantial non-ideal characteristics. Data from different sources may vary significantly in recording conditions, acquisition settings, speaking styles, and annotation conventions, while Tibetan text itself is characterized by complex writing structures and dialectal variation, which often lead to inconsistent symbol usage, irregular non-standard expressions, and unstable speech-text correspondence. These issues jointly weaken data consistency and further affect the model's ability to learn pronunciation patterns, prosodic structures, and speaker characteristics.

    To address this issue, this work treats data construction itself as a key component of low-resource Tibetan speech synthesis and proposes a unified quality enhancement pipeline for low-resource, multi-source, and non-ideal Tibetan data. As shown in Figure 1, the proposed pipeline performs systematic data governance from three aspects, namely audio processing, text normalization, and speech-text consistency verification, with the goal of transforming multi-source heterogeneous raw data into high-quality supervisory data suitable for large-model fine-tuning.
    
    \begin{figure}[h]
		\centering
		\includegraphics[width=1\linewidth]{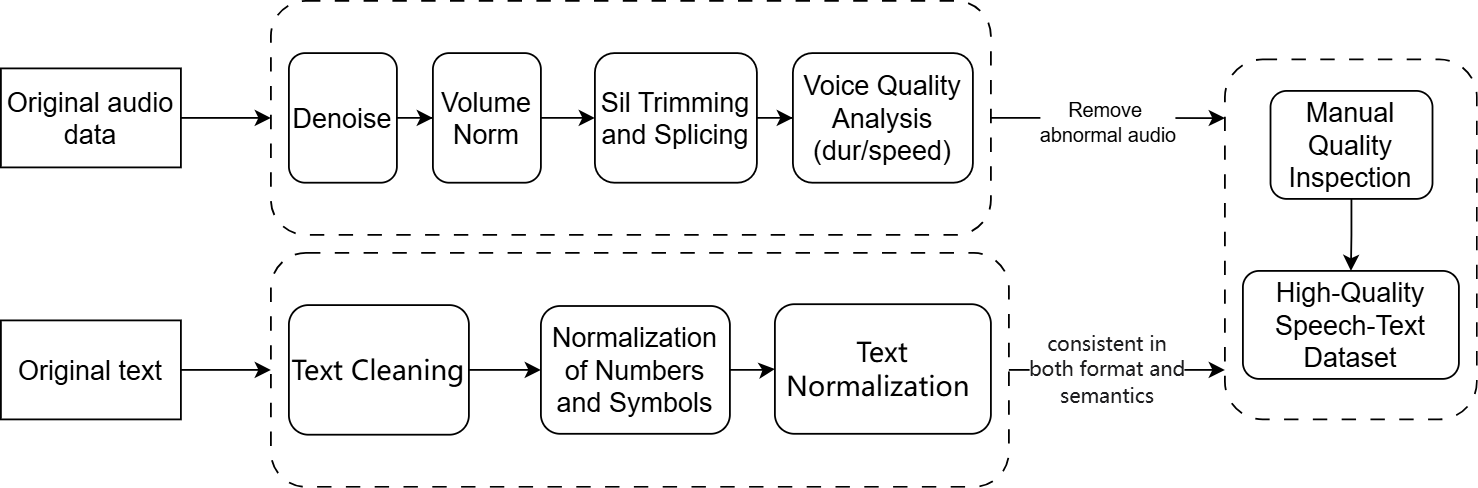} 
		\caption{Unified quality enhancement pipeline for low-resource multi-source Tibetan speech data}
		\label{fig:data-pipeline}
	\end{figure}

    On the audio side, this work mainly addresses common problems in raw recordings, such as noise interference, loudness inconsistency, redundant silence, and abnormal sample distributions, so as to improve the consistency of speech samples in terms of acoustic quality, duration distribution, and trainability. On the text side, a unified normalization mechanism is constructed to handle irregularities commonly found in Tibetan text, including irrelevant symbols, numerical and symbolic expressions, and other non-standard forms, thereby reducing the interference caused by variation in text form during model training.

    On top of this, the proposed pipeline further emphasizes consistency verification of speech-text pairing. In low-resource Tibetan speech synthesis, a critical factor affecting model performance lies not only in unimodal noise, but also in mismatches between speech and text. Therefore, based on automatic processing, this work combines consistency checking with manual quality inspection to further filter and correct speech-text pairs, ultimately constructing a high-quality speech-text parallel dataset.

    Overall, the proposed quality enhancement pipeline is not merely a simple data-cleaning procedure, but a systematic data governance method designed for low-resource, multi-source Tibetan speech synthesis. Centered on audio quality control, text normalization, and speech-text consistency verification, it effectively alleviates the widespread problems of noise, irregular expressions, and pairing errors in raw corpora, and provides a reliable data foundation for subsequent text representation adaptation and cross-lingual transfer modeling.

\subsection{Overall System Architecture}\label{sec:Dialect-ASR-Evaluation}
  
    Based on the previously introduced data quality enhancement pipeline and technical route analysis, this work constructs a modular system architecture for low-resource Tibetan speech synthesis, as illustrated in Fig. 2. The overall framework consists of three major components: lightweight text preprocessing, Tibetan-oriented text representation and tokenizer adaptation, and a speech generation and cross-lingual adaptive training module built upon the large speech synthesis model developed by Xingchen AGI Lab.

    Specifically, the input text first undergoes necessary normalization to reduce the interference caused by non-standard expressions during training and inference. It is then mapped into discrete token sequences by an adapted tokenizer that better matches Tibetan linguistic characteristics. Finally, the token sequence is fed into a speech synthesis module based on the Xingchen large speech model. By leveraging the cross-lingual representation and generation capabilities learned by the Xingchen model from large-scale Chinese-English mixed speech data and multi-dialect speech data, this work further conducts cross-lingual adaptive fine-tuning with high-quality Tibetan speech-text parallel data, thereby enabling stable and natural Tibetan speech synthesis under low-resource conditions.

    On this basis, the following subsections further describe the key modeling components of the proposed system, namely Tibetan-oriented text representation and tokenizer adaptation, as well as the cross-lingual adaptive training strategy.

    \begin{figure}[h]
		\centering
		\includegraphics[width=1\linewidth]{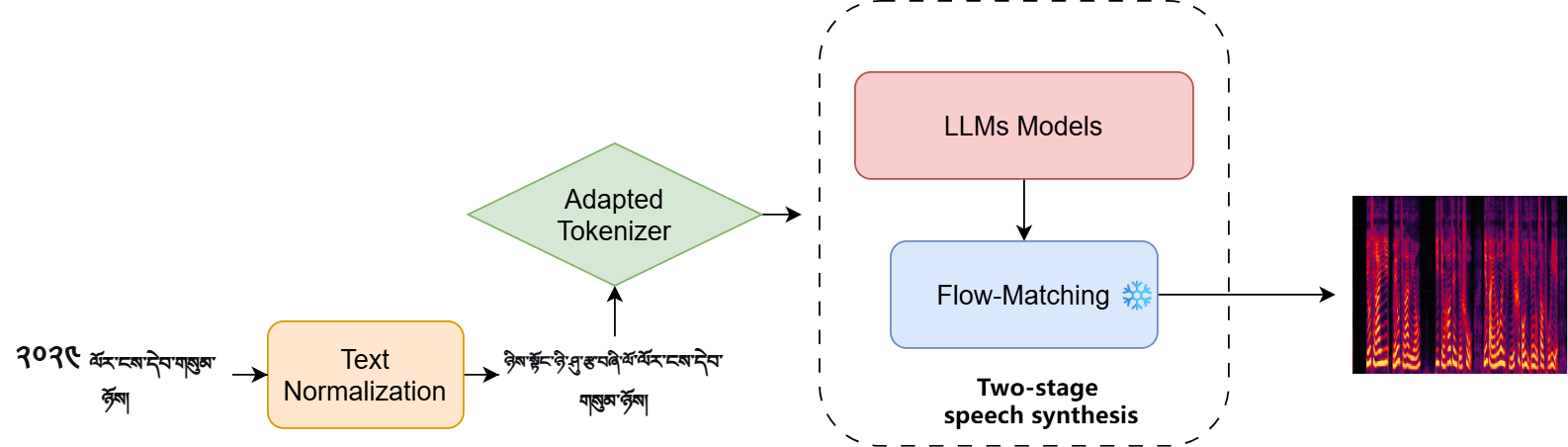} 
		\caption{Overall architecture of the TTS system}
		\label{fig:tts-architecture}
	\end{figure}

\subsection{Tibetan-Oriented Text Representation and Tokenizer Adaptation}

  After data quality enhancement, another key factor affecting model performance under low-resource conditions is how to organize the text input in a way that better matches Tibetan linguistic characteristics. For a speech synthesis system built on the Xingchen large speech model, the input text typically needs to be tokenized before being mapped into a sequence of discrete tokens for subsequent encoding and generation. However, the default tokenization strategies adopted by general pretrained models are mostly designed for high-resource languages such as Chinese and English. Their segmentation units do not fully align with the structure of Tibetan text, and direct application to Tibetan scenarios may introduce representational redundancy, excessively long sequences, and alignment ambiguity, thereby increasing the difficulty of learning the mapping from text to speech.

  More specifically, when processing Tibetan text, the original backbone model typically adopts a fine-grained sub-character or subword segmentation strategy. Although this strategy provides strong generality, it often leads to longer input sequences and increased alignment ambiguity in Tibetan scenarios. In addition, its representation units do not fully match the syllable-centered structure of Tibetan, making it difficult to fully exploit Tibetan phonological priors and thus affecting modeling efficiency and stability under low-resource conditions.

  To address this issue, this work proposes a Tibetan-oriented text representation and tokenizer adaptation method, with the goal of making the input representation better aligned with Tibetan natural linguistic units while preserving as much of the original pretrained model capability as possible. Specifically, two adaptation strategies are explored, as illustrated in Fig. 3.
  
  \begin{figure}[h]
    \centering
	    \includegraphics[width=0.6\linewidth]{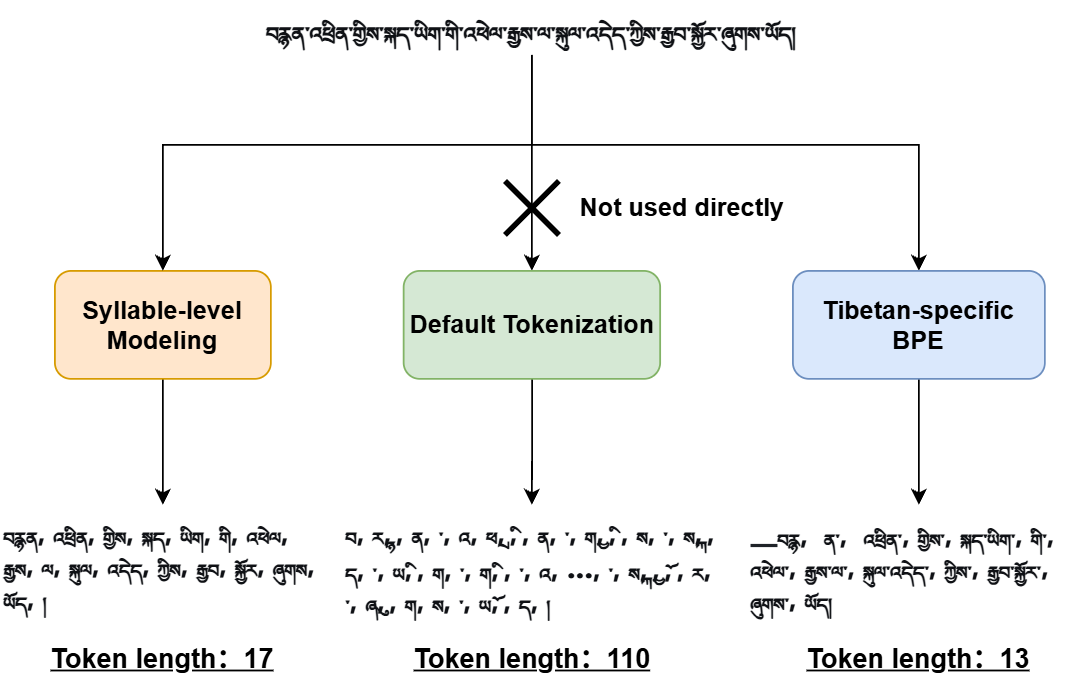}
		\caption{Illustration of Tibetan-oriented text representation and tokenizer adaptation}
		\label{fig:tokenizer-adaptation}
  \end{figure}

  The first strategy is syllable-level modeling. In this strategy, Tibetan syllables or individual characters are treated as the basic modeling units, with explicit separators used to distinguish them in the input sequence. Compared with the default fine-grained segmentation strategy, syllable-level representation more naturally corresponds to Tibetan pronunciation structures and can reduce sequence redundancy and alignment ambiguity to some extent, thereby improving training efficiency and pronunciation stability in low-resource scenarios.

  The second strategy is BPE tokenizer replacement trained on Tibetan corpora. Since the vocabulary distribution of the pretrained tokenizer may not sufficiently cover Tibetan textual characteristics, this work further explores replacing the original tokenizer with a Tibetan-specific BPE tokenizer trained on Tibetan corpora. This strategy makes the statistical distribution of the input representation more consistent with actual Tibetan usage. To some extent, it balances representation compression and preservation of linguistic characteristics, thereby improving text encoding efficiency and enhancing the model's adaptability to Tibetan input.

  It should be noted that the proposed tokenizer adaptation method does not attempt to reconstruct the backbone model itself. Instead, it follows the principle of lightweight yet effective adaptation by performing targeted optimization at the input representation level. Overall, this method provides a more stable and efficient text representation basis for subsequent cross-lingual adaptive training, and serves as an important component for improving the accuracy and stability of low-resource Tibetan speech synthesis.

\subsection{Cross-Lingual Adaptive Training Strategy for Low-Resource Scenarios}

After obtaining relatively high-quality speech-text parallel data and completing Tibetan-oriented text representation adaptation, the core problem in building a low-resource Tibetan speech synthesis system is how to efficiently activate the cross-lingual capability of a pretrained large model under limited supervision. Unlike high-resource languages, Tibetan cannot rely on large-scale, highly consistent task-specific data to train a high-performance speech synthesis model from scratch. Therefore, a more feasible technical route is to fully leverage the universal speech representation capability learned by a pretrained large model from high-resource languages and transfer it to Tibetan and its dialect scenarios through targeted parameter adaptation.

Based on this idea, this work proposes a cross-lingual adaptive training strategy for low-resource scenarios. The strategy takes the large speech synthesis model developed by Xingchen AGI
Lab as the pretrained backbone and performs supervised fine-tuning with Tibetan speech-text parallel data while retaining its general speech modeling and generation capability, so that the model can gradually learn the distributional characteristics of the target language in terms of pronunciation patterns, prosodic organization, and speaker traits. Compared with training entirely from scratch, this approach significantly reduces dependence on target-language data scale and yields more stable training performance under limited data conditions.

The core idea adopted in this work is ``cross-lingual transfer + lightweight adaptation.'' Cross-lingual transfer mainly relies on the universal representations learned by the large speech synthesis model developed by Xingchen AGI Lab from Chinese-English mixed speech data and multi-dialect speech data, enabling the model to retain strong prior capability for speech generation when applied to Tibetan as a low-resource language. Lightweight adaptation, in contrast, emphasizes updating only a limited portion of the model parameters without largely disrupting the original model capability, so that the model can better capture the specific pronunciation rules, rhythmic patterns, and speaker-style characteristics of Tibetan dialects. This training scheme not only improves data efficiency under limited-resource conditions, but also contributes to more stable training and synthesis performance.

Overall, the proposed cross-lingual adaptive training strategy does not rely on large-scale Tibetan-specific data, nor does it require building a complex end-to-end model from scratch. Instead, it achieves effective transfer modeling for Tibetan and its major dialects through targeted linguistic adaptation of the Xingchen Large Speech Model. Together with the previously introduced data quality enhancement and text representation adaptation methods, this work ultimately forms a complete technical framework for low-resource Tibetan speech synthesis, providing a methodological basis for future system optimization and expansion.

\section{Preliminary Evaluation and Usability Validation}
  This work conducts a preliminary evaluation of the proposed low-resource Tibetan speech synthesis method, with a particular focus on the impact of tokenizer adaptation on system stability and synthesis quality. Experimental results show that, even under conditions of limited speech data and weak front-end linguistic support, the proposed Tibetan text-to-speech system can still operate stably and generate synthetic speech with high naturalness and intelligibility, thereby providing an initial verification of its feasibility and practical value in low-resource scenarios.

  From the perspective of system stability, the constructed model demonstrates good synthesis consistency across different text lengths and domains, without exhibiting common problems in low-resource speech synthesis tasks such as repetition, omission, or sequence abnormalities. This indicates that targeted adaptation of the tokenization strategy can effectively alleviate the mismatch between text representation and acoustic modeling without relying on complex hand-crafted linguistic rules, thereby improving the robustness of the system in practical application scenarios.

  The subjective listening test results further support the above findings. As shown in Table 1, in MOS evaluations conducted by 10 native Tibetan speakers, the average scores of the syllable-level modeling and BPE-based modeling systems are 4.28 and 4.35, respectively. Both methods achieve relatively high naturalness and intelligibility, indicating that, under the current resource conditions, the Tibetan speech synthesis system built with the proposed method can effectively learn Tibetan pronunciation characteristics and prosodic structures. Among the two methods, BPE-based modeling performs slightly better in perceptual naturalness, suggesting that more flexible subword-level representations may help improve speech fluency to some extent. However, the small difference between the two also indicates that the system shows good adaptability to different tokenization strategies.

    \begin{table}[h]
    \centering
    \caption{Subjective Evaluation of Tibetan TTS Systems with Different Tokenization Strategies}
    \label{tab:mos_results}
    \begin{tabular}{l c c}
    \hline
    \textbf{Tokenization Strategy} & \textbf{MOS} & \textbf{Syllable Accuracy(\%)} \\
    \hline
    X-API & 3.74  & 93.8 \\
    Syllable-based & 4.28  &  97.6 \\
    BPE-based & 4.35  & 96.6  \\
    \hline
    \end{tabular}
    \end{table}

  In addition to subjective evaluation, this work also conducts syllable-level pronunciation accuracy evaluation. As shown in Table 1, the syllable-level modeling method achieves a pronunciation accuracy of 97.6\%, while the BPE-based modeling method reaches 96.6\%. These results indicate that syllable-level modeling has a slight advantage in pronunciation precision. One possible reason is that, under low-resource conditions, some subword units produced by BPE segmentation may be unstable, and their linguistic units do not always correspond strictly to actual pronunciation units, which can affect the consistency between text representation and acoustic realization. In contrast, syllable-level segmentation aligns more directly with the basic pronunciation units of Tibetan, and therefore shows certain advantages in pronunciation control and acoustic modeling stability.

  To further verify the effectiveness of the proposed method, this work introduces a commercial Tibetan speech synthesis interface provided by a leading domestic speech technology company as a comparison system, denoted as X-API. Under the same evaluation setting, X-API achieves a MOS score of 3.74 and a pronunciation accuracy of 93.8\%, both lower than those of the proposed system. This result indicates that, under low-resource conditions, the proposed method shows competitive performance in both speech naturalness and pronunciation accuracy, and also indirectly validates the effectiveness of the proposed data governance, tokenizer adaptation, and cross-lingual transfer strategies.

  A joint analysis of the subjective and objective results reveals a certain complementarity between different tokenization strategies: BPE-based modeling has a slight advantage in perceptual naturalness, whereas syllable-level modeling performs better in pronunciation accuracy. This suggests that, in low-resource Tibetan speech synthesis, there may be a trade-off between speech fluency and phonological precision. Relatively speaking, subword-level representations are more beneficial for improving overall fluency and generalization, while syllable-level representations are more advantageous for achieving stable and accurate pronunciation control.

  Overall, the above experimental results indicate that the proposed technical route of ``data quality enhancement + text representation adaptation + cross-lingual adaptive training'' enables highly usable Tibetan speech synthesis under relatively low-resource conditions. Compared with traditional front-end solutions that rely heavily on hand-crafted linguistic rules, the proposed method maintains good synthesis quality while reducing system design and maintenance complexity. This is particularly important for low-resource languages such as Tibetan, which often face practical challenges including insufficient language resources, large discrepancies between written forms and spoken pronunciation, and weak supporting toolchains.

\section{Future Work}

This work presents a practical Tibetan text-to-speech system that can operate under low-resource conditions. Built upon a self-developed model and combined with customized Tibetan text normalization and data processing pipelines, the system demonstrates good stability and synthesis quality in both objective evaluations and native-speaker subjective assessments, and is capable of generating natural, clear, and intelligible speech for the \"{U}-Tsang dialect. The results indicate that the proposed method is feasible for low-resource Tibetan speech synthesis and further validate the effectiveness of a large-model-based framework for Tibetan TTS.

Future work will mainly focus on several directions. First, the system will be extended to support additional major Tibetan dialects, and on this basis, a unified multi-dialect Tibetan speech synthesis framework will be further explored. Compared with building separate models for different dialects, a unified multi-dialect model is expected to make better use of the shared linguistic and acoustic regularities across dialects, thereby improving cross-dialect transferability, generalization ability, and data efficiency. Second, the evaluation framework will be further refined. Beyond naturalness and pronunciation accuracy, more comprehensive analyses will be conducted on prosodic expressiveness, emotional variation, long-form stability, and adaptability to real-world scenarios. Third, by combining higher-quality data resources with more efficient model adaptation methods, future work will continue to improve modeling capability under extremely low-resource conditions, while further enhancing zero-shot voice cloning and cross-speaker generalization.

From an application perspective, the proposed framework has strong scalability and practical deployment potential. With the gradual expansion of dialect coverage, the continued development of a unified multi-dialect model, and further optimization of overall system performance, the system is expected to be applied to a wide range of scenarios, including education, broadcasting, public services, and customer service, and may provide a useful reference for speech technology development in Tibetan and other low-resource languages.

\bibliographystyle{IEEEtran}
\bibliography{ref}
\end{document}